\documentclass[aps,prl,twocolumn,superscriptaddress,floatfix]{revtex4}

\usepackage{graphicx}
\usepackage{dcolumn}
\usepackage{bm}
\usepackage{amsmath}
\usepackage{amssymb}
\usepackage{color}

\begin{document}

\renewcommand{\vec}[1]{\mbox{\boldmath $#1$}}


\title{Emergent units of itinerant spin-state excitations in LaCoO$_3$}


\author{K. Tomiyasu}
\email[Electronic address: ]{tomiyasu@tohoku.ac.jp}
\affiliation{Department of Physics, Tohoku University, Aoba, Sendai 980-8578, Japan}
\author{T. Nomura}
\email[Electronic address: ]{nomurat@spring8.or.jp}
\affiliation{National Institutes for Quantum and Radiological Science and Technology, SPring-8, Sayo, Hyogo 679-5148, Japan}
\author{Y. Kobayashi}
\affiliation{Department of Physics, Tokyo Medical University, Shinjuku, Tokyo 160-8402, Japan}
\author{S. Ishihara}
\affiliation{Department of Physics, Tohoku University, Aoba, Sendai 980-8578, Japan}
\author{S. Ohira-Kawamura}
\affiliation{J-PARC Center, Japan Atomic Energy Agency, Tokai, Ibaraki 319-1106, Japan}
\author{M. Kofu}
\affiliation{J-PARC Center, Japan Atomic Energy Agency, Tokai, Ibaraki 319-1106, Japan}



\date{\today}

\begin{abstract}
Spin crossover is expected to enrich unusual physical states in various types of condensed matter. Through inelastic neutron scattering, we study the spin-state excitations in the canonical and advanced platform, LaCoO$_3$, and reveal that the spatial correlation robustly maintains the seven-Co-site size below 300 K and the internal Co-$d$ electrons are spatially delocalized. By combining theoretical calculations, this dynamical short-range order is identified as a new collective unit for describing spin-state with dual spin-state nature beyond the conventional one-Co-site classification. 
\end{abstract}


\maketitle

%
%
The clarification of collective behaviors in many-body systems is a major issue in physics. The interplay between spin, orbit, charge, and lattice governs many physical properties in correlated electron systems. In particular, intriguing critical states and huge responses to external stimuli emerge near the boundary between itinerancy and localization and between magnetism and non-magnetism. The example includes an insulator-metal transition, unconventional superconductivity, high-rank symmetry ordering, colossal magnetoresistance, and spin crossover, also known as spin-state transition. 

The spin-state degree-of-freedom is a unique spin-orbital composite parameter. Perovskite LaCoO$_3$ (nominally Co$^{3+}$: $d^6$) provides its prototype and advanced platform, as this material is considered very proximate to the boundary between the low-spin (LS: $S = 0$, $t_{2g}^{6}$), high-spin (HS: $S = 2$, $t_{2g}^{4}$$e_{g}^{2}$), and intermediate-spin (IS: $S = 1$, $t_{2g}^{5}$$e_{g}^{1}$ with active orbital degree of freedom) states. Starting from the ground insulating nonmagnetic (NM) LS state in the low-temperature (LT) range below $T_{\rm NM} \approx 30$ K, as the temperature increases, magnetic spin-state excitations are thermally activated, resulting in the first spin crossover to a semiconductive paramagnetic state~\cite{Heikes_1964}. Then, the system enters the second spin crossover with an insulator-to-metal transition in the high-temperature (HT) range above $T_{\rm IM} \approx 530$ K, at which the specific heat exhibits a maximum~\cite{Stolen_1997}. Furthermore, the thin-film fabrication, surface and interface of bulk, and slight oxygen defect break the LS non-magnetism and generate the ferromagnetism~\cite{Fuchs_2007, Fujioka_2013, Fujioka_2015, Durand_2013, Kaminsky_2018, Giblin_2009, EI-Khatib_2015}. In theory, LaCoO$_3$ also virtually undergoes the quantum spin-state exciton condensation, which is analogous to the mechanism of superconductivity~\cite{Afonso_2017}. 

\begin{figure}[h!tbp]
\begin{center}
\includegraphics[width=0.81\linewidth, keepaspectratio]{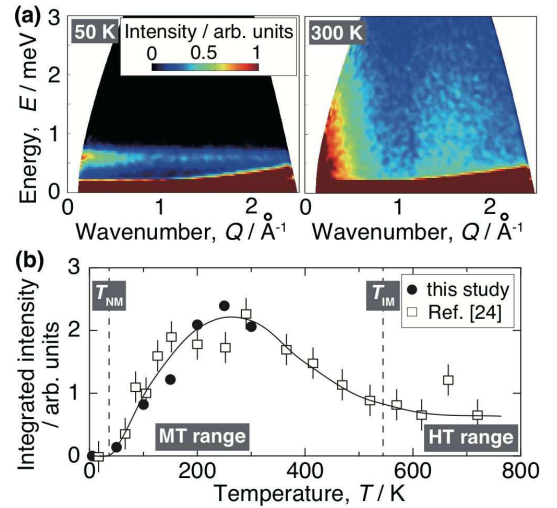}
\end{center}
\caption{\label{fig:pwdav} (Color online)
Orientational-averaging powder-like neutron data. (a) Intensity maps measured at 50 K (left) and 300 K (right) in the $(Q,E)$ space. (b) Temperature dependence of $Q$--$E$ integrated intensity after subtracting the background. The curve is a guide for the eye. The statistical errors of the solid circles are smaller than the symbol size. 
}
\end{figure}

In the middle-temperature (MT) range between $T_{\rm NM}$ and $T_{\rm IM}$, it has been controversial for a long time, whether the thermally excited states are HS ($J_{\rm eff} = 1$, $g \sim 3.4$, $E \sim 13$ meV) \cite{Noguchi_2002, Haverkort_2006, Podlesnyak_2006, Tomiyasu_2017, Shimizu_2017}, IS~\cite{Korotin_1996, Saitoh_1997, Asai_1998}, or both~\cite{Asai_1998, Radaelli_2002, Sato_2009}. These atomic pictures are effective even in the presence of strong Co--O covalent bonding 
~\cite{Krapek_2012}, and the HS description is currently believed to be the closest to the truth. However, there still exists a major problem, namely, although the HS--HS interaction is supposed to be antiferromagnetic from both the Goodenough-Kanamori rule and magnetic susceptibility data~\cite{Kyomen_2003}, ferromagnetic short-range correlation is observed in the spin-state excitations in a wide MT range by neutron scattering, which is the prime tool for studying spin~\cite{Asai_1989, Asai_1994, Phelan_2006_a}. This correlation disappears in the LT range, but as the temperature increases, it begins to grow as gapped excitations at 0.6 meV from $T_{\rm NM}$ and becomes gapless (quasielastic) excitations above 100 K~\cite{Podlesnyak_2006, Phelan_2006_a}. The intensity shows a broad maximum from 200 to 300 K and disappears toward $T_{\rm IM}$~\cite{Asai_1989, Asai_1994}. For clarity, we show our neutron data measured below 300 K in Figs.~\ref{fig:pwdav}(a) and \ref{fig:pwdav}(b) together with the Ref.~[\onlinecite{Asai_1994}] data. 
Thus, these ferromagnetic short-range excitations characterize the MT range.  

%

In this study, to resolve the controversy regarding the HS-or-IS problem and supporting experimental facts in the MT range, we investigate the spatial spin distribution of the spin-state excitations from 5 to 300 K with the single-crystal inelastic neutron scattering technique. Furthermore, by combining theoretical calculations, we clarify that the dynamical short-range order exhibits the HS and IS dual nature as a collective unit. 

%
%
{\it Experiments}.-- 
The neutron scattering experiments were performed on the chopper spectrometer AMATERAS (BL14) at the MLF of the J-PARC spallation neutron source (Japan)~\cite{Nakajima_2011}. 
The incident energy ($E_{\rm i}$) was set to 4.7 and 15 meV, and the $E$ resolution under elastic conditions was approximately 2.3 and 3.6{\%} to $E_{\rm i}$, respectively. The speed of the main disk chopper was fixed at 300 Hz. The data were obtained by the UTSUSEMI software provided by the MLF~\cite{Inamura_2013}. A single-crystal sample with a length of 40 mm and a diameter of 6 mm was grown in the O$_2$ gas flow by the floating-zone method and was annealed at 750 $^{\circ}$C for 3 hours in the O$_2$ gas flow. The crystal rod was mounted under a cold head in a He closed-cycle refrigerator. 

\begin{figure}[t]
\begin{center}
\includegraphics[width=0.81\linewidth, keepaspectratio]{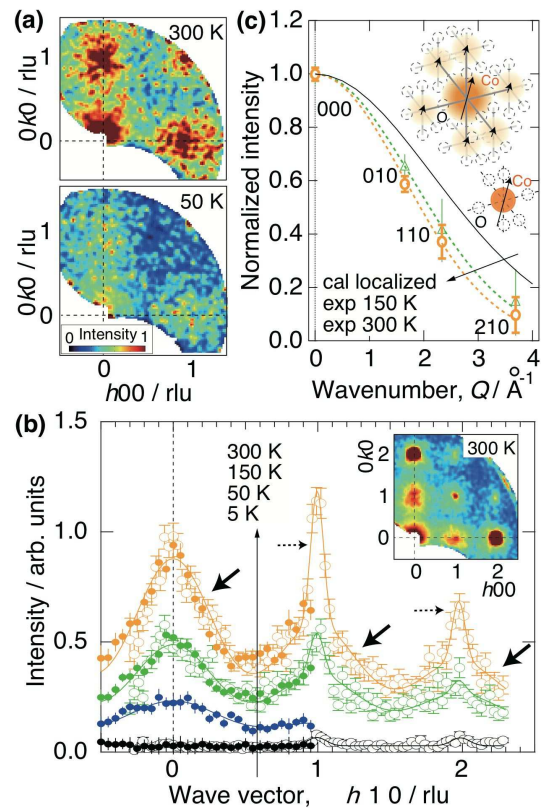}
\end{center}
\caption{\label{fig:sx} (Color online)
Single-crystal neutron data. 
(a) Intensity maps measured at 300 K (upper) and 5 K (lower) in the $(hk0)$ zone for $E_{\rm i}=4.7$ meV. The integration range is $l = -0.2$ to 0.2 reciprocal lattice unit (rlu) and $E = 0.5$ to 0.7 meV. 
(b) Cuts on the $h10$ line. Solid symbols represent the data obtained by integrating $k=-0.1$ to 0.1 rlu in (a). Open symbols represent the data, measured for $E_{\rm i}=15$ meV and obtained by integrating $k= -0.1$ to 0.1 rlu, $l = -0.2$ to 0.2 rlu, and $E = 1.5$ to 3.5 meV. The curves show the results of least-square fitting using multiple Gaussians. The diagonal solid arrows indicate diffuse scattering, while the horizontal dotted arrows indicate sharp scattering by acoustic phonons. The inset shows the intensity map measured for $E_{\rm i}=15$ meV. 
(c) $Q$ dependence of diffuse scattering intensity extracted by the fitting in (b). The solid line denotes the square of the localized magnetic form factor~\cite{Watson_1961}. The broken lines denote the guide for the eye. The upper inset illustrates the seven-Co-site spin-state unit, in which the colored balls indicate the delocalized spin clouds and the arrows denote spins. 
The spin ratio of the nearest neighbor Co to the center Co is approximately 0.1--0.2.
The lower inset illustrates the one-Co-site localized spin-state for comparison. } 
\end{figure}

%
%
{\it Results.}-- 
Hereinafter, the pseudo-cubic notation is used, while the exceptional usage of rhombohedral notation is denoted by the subscript r.
Figure~\ref{fig:sx}(a) shows the maps of inelastic neutron scattering intensity measured in the $(hk0)$ scattering plane for $E_{\rm i}=4.7$ meV. The diffuse scattering is observed around the fundamental reciprocal lattice points of 000, 100, 010, and 110 in the wide temperature range from 50 to 300 K. This unambiguously indicates that the spin excitations are ferromagnetic. In addition, it is newly observed that the distribution pattern is overall temperature-independent, while the magnitude of intensity largely changes with the temperature. 

The left side in Fig.~\ref{fig:sx}(b) (solid symbols) shows the line cuts obtained at several temperatures. The half-width at half maximum (HWHM) of the diffuse scattering, $\Delta Q_{1}$, is estimated as 0.39 and 0.33 rlu at 50 and 300 K, respectively. These correspond to the short correlation length, $\xi_{1} = 0.60d_{\rm Co-Co}$ and $0.70d_{\rm Co-Co}$, where $d_{\rm Co-Co}$ denotes the nearest neighbor Co--Co distance ($\approx 3.8$ {\AA}). 

To cover the higher $Q$ range, we also use the data measured for $E_{\rm i}=15$ meV. As shown by the open symbols in Fig.~\ref{fig:sx}(b), whereas the 0.6-meV gap at 50 K cannot be resolved because of the lower energy resolution, these data are useful for studying the gapless excitations at 150 and 300 K and the consistency with the data measured for $E_{\rm i}=4.7$ meV is confirmed. Furthermore, the scattering by acoustic phonons arising from the fundamental Bragg reflections is distinctively sharp, as typically seen at the 110 point. Therefore, the intensity of diffuse scattering at several $\vec{Q}=hkl$ points can be extracted by the least-square fitting using multiple Gaussians, in which the single diffuse width ($\Delta Q_{1}$) and the single sharp width are commonly used for all the $hkl$ points and the two $E_{\rm i}$ data. The resultant curves are also drawn in Fig.~\ref{fig:sx}(b), indicating a satisfactory fit with the experimental data.

The normalized integrated intensity of diffuse scattering thus extracted is plotted as a function of $Q=|\vec{Q}|$, $I(Q)$, in Fig.~\ref{fig:sx}(c). $I(Q)$ decreases with increasing $Q$, confirming that the diffuse scattering is magnetic in origin, in accordance with the magnetic form factor. 
Furthermore, the degree of decrease is remarkably faster than the square of the theoretical localized magnetic form factor (solid line)~\cite{Watson_1961}. By the uncertainty principle, this indicates that the spatial spin cloud is fairly delocalized around each Co site. For clarity, we define the effective radius of the cloud, $\xi_{2}$, by taking the inverse of the HWHM of $I(Q)$, $\Delta Q_{2}$, and scaling the localized value to the Shannon ionic radius, $\xi_{2}({\rm local}) \equiv \alpha \cdot \Delta Q_{2}({\rm local})^{-1} \equiv r({\rm Co^{3+{\rm (HS)}}}) =  0.61$ {\AA}~\cite{Shannon_1976}, where $\alpha$ denotes the scaling constant. The obtained values are $\xi_{2}({\rm 150K}) = 0.77$ {\AA} and $\xi_{2}({\rm 300K}) = 0.84$ {\AA}, which are fairly larger than the hard radius of Co defined by $r_{\rm hard} \equiv d_{\rm Co-O} - r({\rm O^{2-}}) = 0.59$ {\AA}~\cite{Radaelli_2002, Shannon_1976}, that is, $\xi_{2}({\rm local}) \approx r_{\rm hard}$ but clearly $\xi_{2}({\rm 300K}) > r_{\rm hard}$. 
This indicates the substantial spatial overlap of the Co--O electron cloud, which will lead to the strong $p$--$d$ hybridization thus far discussed~\cite{Saitoh_1997, Haverkort_2006, Kobayashi_2015, Tomiyasu_2017}.
We also remark that this type of delocalized magnetic form factor is rare but is observed in some organic radical $\pi^{*}$ molecules, cuprates related to superconductivity, and a spin-orbit frustrated heavy-fermion metal~\cite{Zheludev_1994_b, Walters_2009, Tomiyasu_2014}. 

In this way, we found the following characteristics for the short-range ferromagnetic excitations. 1) The spatial correlation is approximately the seven-Co-site size and robustly sustains below 300 K. 2) The internal Co-$d$ electrons are fairly delocalized. 

\begin{figure}[t]
\begin{center}
\includegraphics[width=1.0\linewidth]{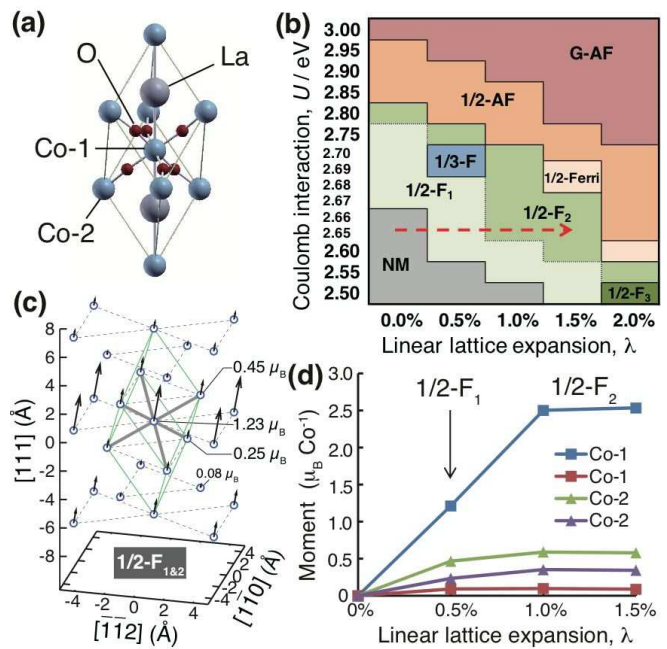}
\end{center}
\caption{
\label{fig:cal} (Color online)
(a) Rhombohedral cell containing two Co and six O atoms. 
(b) Calculated phase diagram. The $U$ range near the NM edge is magnified. 
F, Ferri, and AF denote ferromagnetic, ferrimagnetic, and antiferromagnetic states, respectively; 
1/2 and 1/3 the amplitude-modulated states described by $\vec{Q}_{1/2}$ and $\vec{Q}_{1/3}$, respectively; 
G the G-type. 
The red dotted arrow denotes the expected path for thermal spin crossover in bulk. 
(c) Real-space spin structure. The magnetic moment values obtained for 1/2-F$_1$ are also written. The green rhombohedron represents the cell, while the thick gray lines show the seven-Co-site unit. The arrows denote the spin moments. The dotted triangular planes indicate the (111) planes. 
(d) Evolution of spin moments calculated along the arrow in (b). }
\end{figure}

%
%
{\it Approach by theoretical calculations.}--
Considering the experimental fact that the lattice volume significantly increases with the temperature increasing toward $T_{\rm IM}$~\cite{Asai_1998}, we started with first-principle band calculations for the unexpanded and expanded lattices by using the WIEN2k code~\cite{Blaha_2017}. As the unexpanded lattice, we used the structural parameters determined at 10 K by neutron diffraction~\cite{Asai_1998, Radaelli_2002}. As the expanded lattice, we enlarged the lattice constants while keeping the fractional coordinates in the $R{\bar 3}c$ space group for simplicity. From the bands near the Fermi energy, we constructed 28 maximally localized Wannier orbitals, consisting of 10 Co-$d$ and 18 O-$p$ orbitals, by using the rhombohedral cell [Fig.~\ref{fig:cal}(a)]. 
Thus, we obtained the corresponding tight-binding model with the aid of the wannier90 code~\cite{Mostofi_2008, SM} and the 28-orbital Hubbard model. For the Coulomb interactions  ($U$ and $U'$) and the Hund coupling ($J_{\rm H}$) at each Co site, we retained the relations, $U = U'+2J_{\rm H}$ and $J_{\rm H}=0.2U$. 
Then, to analyze the magnetic properties, the Hartree-Fock mean-field approximation was used~\cite{SM}. 
We examined the following cases;  
(i) $\vec{q}_1 = (0,0,0) \equiv \vec{Q}_{0}$; 
(ii) $\vec{q}_1 = \vec{Q}_{0}$ and $\vec{q}_2 = (1/2,1/2,1/2)_{\rm r} \equiv \vec{Q}_{1/2}$;  
(iii) $\vec{q}_1 = \vec{Q}_{0}$ and $\vec{q}_{2,3} = \pm (1/3,1/3,1/3)_{\rm r} \equiv \pm \vec{Q}_{1/3}$, 
where $\vec{q}_s$ denotes the ordering wave vector. 
The modulating $\vec{Q}_{1/2}$ and $\vec{Q}_{1/3}$ contain $(1/4,1/4,1/4)$ and $(1/6,1/6,1/6)$ observed for thin-film LaCoO$_3$, respectively~\cite{Fujioka_2015}. 
Further details are summarized in the Supplementary Material~\cite{SM}. 

Figure~\ref{fig:cal}(b) shows the phase diagram near the nonmagnetic-to-magnetic boundary found in the space of linear lattice expansion $\lambda$ and Coulomb $U$, where $\lambda = \Delta L/L$, $L$ denotes the linear dimension of the lattice, and $\Delta L$ its variation. 
The boundary exists around $\lambda = 0.5${\%} and several magnetic phases sensitively vary in this narrow $(\lambda, U)$ range. Further, the modulated order appears before the well-known G-type antiferromagnetic and uniform ferromagnetic order normally expected for perovskites. These features are consistent with the thin-film experiments~\cite{Fuchs_2007, Fujioka_2013, Fujioka_2015, Sterbinsky_2018}. 
%

%

Bulk LaCoO$_3$ exhibits the two steps of thermal spin crossover, in which typically $\lambda = 0.5${\%} at 300 K (MT range) and $\lambda = 1.2${\%} at 600 K (HT range)~\cite{Radaelli_2002}. Therefore, we choose the value of $U \approx 2.65$ eV, as shown by the red arrow in Fig.~\ref{fig:cal}(b); the NM state changes to the 1/2-F$_1$ at $\lambda = 0.5${\%} followed by the 1/2-F$_2$ at $\lambda = 1.0${\%}. 
Those spin structures are described by $\vec{Q}_{0}$ and $\vec{Q}_{1/2}$ and are nearly homothetic to each other, in which the ferromagnetic (111) sheets are stacked along the [111] direction under modulation of the spin amplitude, as depicted in Fig.~\ref{fig:cal}(c). 
However, the bulk $\lambda$ expansion results from the thermal spin-state excitations, which are inherently dynamical and short-ranged. Furthermore, in general, the mean-field approximation tends to overestimate the ordering. In particular, the 1/2-F$_1$ energy lowers by less than 1 eV per cell from the NM~\cite{SM}. Hence, it is reasonable to moderately consider that the theoretically obtained 1/2-F$_1$ and 1/2-F$_2$ long-range structures are timely and spatially dissipated as the thermal excitations. 
In fact, our neutron scattering exhibits the dynamical seven-Co-site spin structure [the upper inset in Fig.~\ref{fig:sx}(c)], which indeed inheres as the seed of the 1/2-F$_1$ and 1/2-F$_2$ structures, as shown by the thick gray lines in Fig.~\ref{fig:cal}(c). 
The scattering around $\vec{Q}_{1/2}$ is experimentally not detected, which is understood because this seed structure is smaller than its modulation period. In addition, the same spatial spin correlation persists in the HT range [Fig.~\ref{fig:pwdav}(b)], which is also consistent with the homothety of 1/2-F$_1$ and 1/2-F$_2$. 

In this way, we found the amplitude-modulated ferromagnetic order for the expanded lattice. This probably corresponds to the long-range structure observed in thin films, of which the seed structure appears as the thermally dissipated seven-Co-site excitations in bulk.


Finally, we present the $d$-electron configurations at the most spin-polarized Co site in Table~\ref{tbl:econfig}. The changes from NM via 1/2-F$_1$ to 1/2-F$_2$ represent the two steps of spin crossover. The three states accommodate approximately 7 electrons, not 6, suggesting the major contribution of the $d^{7}\underline{L}$ state in the language of CoO$_6$ cluster, where $\underline{L}$ denotes an oxygen hole. The $t_{2g}$ orbital splits into doublet and singlet while the $e_g$ orbital is degenerate. All these facts are in agreement with the previous theoretical and experimental studies~\cite{Korotin_1996, Saitoh_1997, Haverkort_2006, Tomiyasu_2017, Ishikawa_2004, Kobayashi_2005}. 
However, their interpretations split on whether it is HS or IS. Furthermore, even the most polarized spin-moment value is 1.23 $\mu_{\rm B}$ in the 1/2-F$_1$, which is distinctively smaller than both the expected values for the HS and IS states (4 and 2 $\mu_{\rm B}$). 

%
\begin{table}[htbp]
\caption{\label{tbl:econfig} 
Calculated $d$-electron spin and orbital configurations at the most spin-polarized Co site for $U=2.65$ eV. $n$ denotes the total electron filling; $m$ the total spin moment ($\mu_{\rm B}$); $x$, $y$, and $z$ are the pseudo-cubic directions. Bold numbers represent the configurations that mainly change from the NM state. The 1/2-F$_1$ and 1/2-F$_2$ states result from the $t_{2g}$-to-$e_g$ transfers of 0.6 and 1.2 electrons, respectively. }
\begin{ruledtabular}
\begin{tabular}{ll ccccc rr}
State &   & $xy$ & $yz$ & $zx$ & $3z^{2}-r^{2}$ & $x^{2}-y^{2}$ & $n$ & $m$ \\
\hline
NM                                 & $\uparrow$ & 0.96 & 0.96 & 0.96 & 0.36 & 0.36 & 7.20 & 0\\
($\lambda=0.0 \%$)       & $\downarrow$   & 0.96 & 0.96 & 0.96 & 0.36 & 0.36 &  & \\
\hline
1/2-F$_1$                     & $\uparrow$ & 0.99       & 0.98 & 0.98 & {\bf 0.59}  & {\bf 0.59} & 7.03 & 1.23\\
($\lambda=0.5 \%$)       & $\downarrow$   & {\bf 0.38} & 0.95 & 0.95 & 0.31 & 0.31       &  & \\
\hline
1/2-F$_2$                     & $\uparrow$ & 0.99       & 0.99       & 0.98 & {\bf 0.85}  & {\bf 0.85} & 6.82 & 2.50\\
($\lambda=1.0 \%$)       & $\downarrow$  & {\bf 0.33} & {\bf 0.33} & 0.95 & 0.28 & 0.27 &  & \\
\end{tabular}
\end{ruledtabular}
\end{table}
%

%
%
{\it Discussion.}-- Thus, we try to attain insight a step further for the MT spin state beyond the conventional one-Co-site HS-or-IS classification. First, experimentally, nearly the same seven-Co-site correlation sustains in the MT range [Result 1)]. This behavior is as if the seven-Co-site structure object is a new collective robust spin-state unit, or emergent spin-state excitation unit. Furthermore, significant delocalization is observed in this correlation [Result 2)]. This suggests that the $d$-electrons itinerantly support the seven-Co-site collectivity including oxygens, which may be robustly stabilized by the semi-local band formation beyond the hybridization. 

Along this concept of the multi-Co-site spin-state unit, examining the summation of spin moments at seven-Co-sites in the 1/2-F$_1$, we obtain the theoretical value of 3.3 $\mu_{\rm B}$ in total ($1.23 + 3\cdot0.25 + 3\cdot0.45$ [Fig.~\ref{fig:cal}(c)]). This is close to the expected value of the one-Co-site HS model. Furthermore, the $d$-electron configuration is approximately regarded as being midway from NM toward $d^7$-HS ($t_{2g}^{5}e_{g}^{2}$) states by focusing on the majority orbitals. 
On the other hand, the intra-unit spatial spin correlation is amplitude-modulated (seed of antiferromagnetism) and parallel (ferromagnetism), which appears to be both the HS and IS characteristics. Furthermore, our theory indicates that the $t_{2g}$ orbital splits, suggesting that the intra-unit electron system acquires the IS-like Jahn-Teller instability or orbital-ordering tendency owing to the internal itinerant collectivity, which is similar to the Korotin theory~\cite{Korotin_1996}. 

Thus, the apparently conflicting multiple aspects simultaneously accompany this collective spin-state unit. This is where the long-standing controversy originates, in which both the HS and IS characteristics appear depending on the scope of each experiment. 
It will be fruitful to consider many experimental reports on LaCoO$_3$ and relevant systems with this concept. Further discussion is given in the Supplementary Material~\cite{SM}. 

%
%
{\it Conclusions.}---
In the single-crystal inelastic neutron scattering, we found the characteristics of temperature-robust spatial correlation and internal delocalization for the spin-state excitations. By combining the theoretical calculations, we identified these excitations as the new collective spin-state units, where the HS and IS dual nature originate in the MT range. 
This concept of a collective excitation structure object could be exploited as the key for understanding the critical phenomena in various many-body systems and as the generator of multi-functionality used for advanced material design.

\acknowledgments
We thank Dr. Y. Inamura for assisting with the reduction of neutron data. The neutron experiments were performed with the approval of J-PARC (2017A0268 and 2013P0202 (PI: K. Nakajima)). This study was financially supported by MEXT and JSPS KAKENHI (JP18K03503, JP17H06137, and JP15H03692) and by the FRIS Program of interdisciplinary research at Tohoku University.

\appendix
\begin{widetext}
\section{Supplementary Material}

\title{Supplementary Material. \\ 
Emergent units of itinerant spin-state excitations in LaCoO$_3$
}

\section{Details of theoretical calculations}
\subsection{Construction of effective tight-binding model}
The Supplementary Material is also written in the pseudo-cubic notation. The exceptional usage of rhombohedral notation is expressed by the subscript r.

We started with the first-principle band calculations by using WIEN2k~\cite{Blaha_2017} for the NM state, in which the $R\bar{3}c$ structural parameters experimentally determined at $T=10$ K~\cite{Radaelli_2002} were used. The bands obtained near the Fermi level were dominated by the Co-$d$ and O-$p$ orbital states. Then, by using the wannier90 code~\cite{Mostofi_2008}, we generated the maximally localized Wannier functions (MLWFs) with those orbital characteristics and obtained the transfer integrals and the one-particle energy levels for them. Figure~\ref{fig:bands-wanniers}(a) shows the band structures, in which the constructed MLWF results reproduce the first-principle results well. Figure~\ref{fig:bands-wanniers}(b) displays the MLWFs, in which the local [111]$_{\rm local}$ direction is defined to be parallel to the rhombohedral [111]$_r$ direction, and the [100]$_{\rm local}$, [010]$_{\rm local}$, and [001]$_{\rm local}$ axes are approximately parallel to the nearest-neighbor Co--Co directions, {\it i.e.}, the pseudo-cubic principal $x$, $y$, and $z$ axes. The obtained MLWF states are energetically split into the triply degenerate $t_{2g}$ states and the doubly degenerate $e_g$ states and the $10D_q$ value is 0.82 eV. 

%
\begin{figure}[htbp]
\begin{center}
\includegraphics[width=0.50\columnwidth]{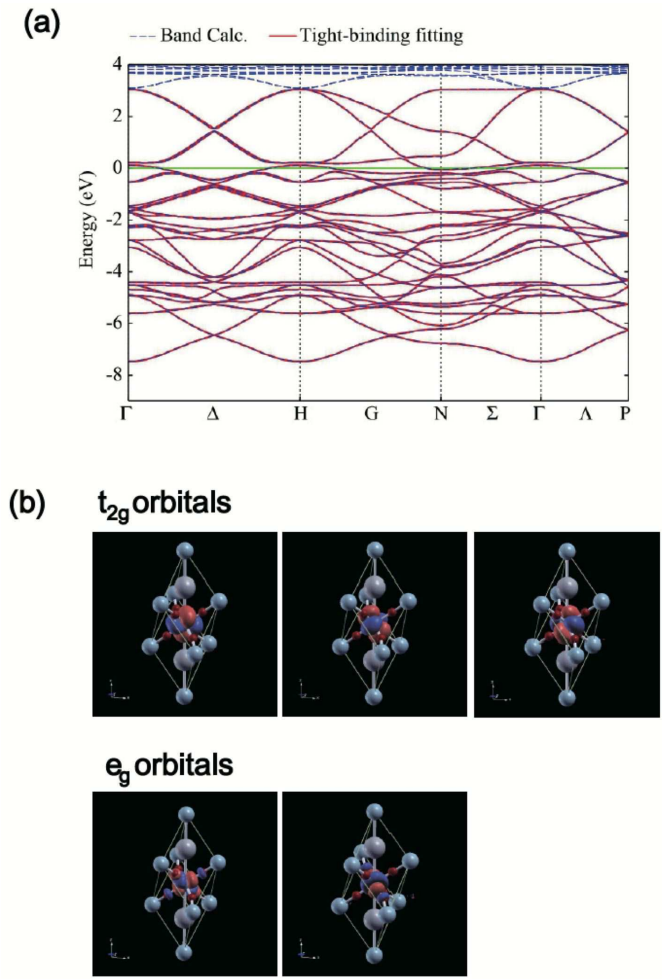}
\end{center}
\caption{
(Color online)
(a) Band fitting. The Fermi energy is set to 0 eV. 
(b) Maximally localized Wannier states at the central Co site. 
They are approximately identical to the $d_{xy}$, $d_{yz}$, $d_{zx}$, $d_{x^{2}-y^{2}}$, and $d_{3z^{2}-r^{2}}$ orbitals. }
\label{fig:bands-wanniers}
\end{figure}
%

\subsection{Hartree-Fock Mean-Field Theory}
\label{SM:HF}

The Hubbard Hamiltonian consists of the non-interacting and interacting parts: 
\begin{equation}
H=H_0 + H'.
\end{equation}
The non-interacting part is given by 
\begin{equation}
H_0= \sum_{i} \sum_{\ell\sigma} \varepsilon_{\ell} 
a^{\dag}_{i\ell\sigma} a_{i\ell\sigma} 
+ \sum_{i,j} \sum_{\ell\ell'\sigma} t_{\ell, \ell'}({\bf r}_i - {\bf r}_j) 
a^{\dag}_{i\ell\sigma} a_{j\ell'\sigma}, 
\label{eq:H_0}
\end{equation}
where $a_{i\ell\sigma}$ and $a^{\dag}_{i\ell\sigma}$ are the electron annihilation and creation operators on the Co-$d$ and O-$p$ states. The one-particle energies $\varepsilon_{\ell}$ and transfer integrals $t_{\ell, \ell'}({\bf r}_i - {\bf r}_j)$ are determined as mentioned above. 
For the interacting part, we take the on-site Coulomb interaction among the Co-$d$ states: 
\begin{eqnarray}
H' &=& \sum_{i} \biggl[ \frac{U}{2}  \sum_{\ell} \sum_{\sigma \neq \sigma'}
d_{i\ell\sigma}^{\dag} d_{i\ell\sigma'}^{\dag} d_{i\ell\sigma'} d_{i\ell\sigma} 
 + \frac{U'}{2} \sum_{\ell\neq\ell'} \sum_{\sigma,\sigma'}
d_{i\ell\sigma}^{\dag} d_{i\ell'\sigma'}^{\dag} d_{i\ell'\sigma'} d_{i\ell\sigma} 
 + \frac{J_{\rm H}}{2} \sum_{\ell\neq\ell'} \sum_{\sigma,\sigma'}
d_{i\ell\sigma}^{\dag} d_{i\ell'\sigma'}^{\dag} d_{i\ell\sigma'} d_{i\ell'\sigma} 
 \biggr], 
\label{eq:H'}
\end{eqnarray}
where $U$, $U'$, and $J_{\rm H}$ are the on-site Coulomb integrals. 

In the Hartree-Fock (HF) mean-field theory, $H$ is approximated as 
\begin{eqnarray}
\label{Eq:HMF}
H_{\rm HF} &=& \sum_{i} \sum_{\ell\sigma} \varepsilon_{\ell} 
a^{\dag}_{i\ell\sigma} a_{i\ell\sigma} + \sum_{i,j} \sum_{\ell\ell'} \sum_\sigma 
t_{\ell,\ell'}({\bf r}_i-{\bf r}_j) a_{i \ell \sigma}^{\dag} a_{i' \ell' \sigma} 
+ \sum_i \sum_{\ell} \biggl[ \frac{U}{2} \langle n_{i\ell} \rangle 
+ \sum_{\ell'(\neq \ell)} \biggl (U'- \frac{J_{\rm H}}{2}  \biggr) \langle n_{i\ell'} \rangle \biggr] n_{i\ell} \nonumber\\
&& - \sum_i \sum_{\ell} \biggl[ \frac{U}{2} \langle {\bf m}_{i\ell} \rangle + \sum_{\ell'(\neq \ell)} 
\frac{J_{\rm H}}{2} \langle {\bf m}_{i\ell'} \rangle \biggr] \cdot {\bf m}_{i\ell} 
- \sum_i \sum_{\ell} \frac{U}{4} \biggl( \langle n_{i\ell} \rangle^2 - | \langle {\bf m}_{i\ell} \rangle |^2 \biggr) \nonumber \\
&& - \sum_i \sum_{\ell \neq \ell'} \frac{U'}{2} \langle n_{i\ell} \rangle \langle n_{i\ell'} \rangle 
+ \sum_i \sum_{\ell \neq \ell'} \frac{J_{\rm H}}{4} \biggl( \langle n_{i\ell} \rangle \langle n_{i\ell'} \rangle 
+ \langle {\bf m}_{i\ell} \rangle \cdot \langle {\bf m}_{i\ell'} \rangle \biggr), 
\end{eqnarray}
where the bracket $\langle X \rangle$ denotes the mean-field value of $X$, and 
\begin{eqnarray}
n_{i\ell} &=& \sum_{\sigma} d_{i\ell\sigma}^{\dag} d_{i\ell\sigma}, \\
{\bf m}_{i\ell} &=& \sum_{\sigma_1\sigma_2} d_{i\ell\sigma_1}^{\dag} 
[{\vec{\sigma}}]_{\sigma_1\sigma_2} d_{i\ell\sigma_2}, 
\end{eqnarray}
where $[{\vec{\sigma}}]$ denotes the Pauli matrix. 
For spatially modulated states, we need to assume that the mean-field values, 
\begin{eqnarray}
\langle n_{i \ell} \rangle     = \sum_{s=1}^{N_s} \langle n_{{\bf q}_s \ell} \rangle \exp(i{\bf q}_s \cdot {\bf r}_i), \\
\langle {\bf m}_{i \ell} \rangle = \sum_{s=1}^{N_s} \langle {\bf m}_{{\bf q}_s \ell} \rangle \exp(i{\bf q}_s \cdot {\bf r}_i), 
\end{eqnarray}
are not zero, where ${\bf q}_s$' denote the ordering wave vectors ($s=1$ to $N_s$). 
The self-consistent equations for the mean fields are described in the momentum ({\bf k}) representation as 
\begin{eqnarray}
\langle n_{{\bf q}_s\ell} \rangle &=& \frac{1}{N_{\bf k}} \sum_{{\bf k},a} \sum_{s'} \sum_{\sigma} 
u_{{\bf q}_{s'}\ell\sigma, a}^*({\bf k}) u_{{\bf q}_{s'}+{\bf q}_s\ell\sigma, a}({\bf k}) n_a({\bf k}), \\
\langle {\bf m}_{{\bf q}_s\ell} \rangle &=& \frac{1}{N_{\bf k}} \sum_{{\bf k},a} \sum_{s'} \sum_{\sigma_1\sigma_2} 
u_{{\bf q}_{s'}\ell\sigma_1, a}^*({\bf k}) [{\vec{\sigma}}]_{\sigma_1 \sigma_2} 
u_{{\bf q}_{s'}+{\bf q}_s\ell\sigma_2, a}({\bf k}) n_a({\bf k}),
\end{eqnarray}
where $u_{{\bf q}_s\ell\sigma, a}({\bf k})$ is the diagonalization matrix of $H_{\rm HF}$ in the momentum representation, 
and $n_a({\bf k})$ is the electron occupation number on diagonalized band $a$ at momentum ${\bf k}$. 
Integrations with respect to ${\bf k}$ were performed by dividing the first Brillouin zone into $N_{\bf k} = 24^3$ meshes, 
and summations with respect to ${\bf k}$ in the above self-consistent equations are performed over the folded Brillouin zone. 

The one-particle energy $\varepsilon_{\ell}$ already includes the following energy shift from the bare one: 
\begin{equation}
\Delta \varepsilon_{\ell} \equiv \frac{U}{2} \langle n_{i\ell} \rangle 
+ \sum_{\ell'(\neq \ell)} \biggl (U'- \frac{J_{\rm H}}{2}  \biggr) \langle n_{i\ell'} \rangle, 
\label{Eq:Delta}
\end{equation}
which is because of the electron Coulomb interaction . Therefore, we determine the bare one-particle energy, 
$\varepsilon_{\ell}^{(0)} \equiv \varepsilon_{\ell} - \Delta \varepsilon_{\ell}$, 
where $\Delta \varepsilon_{\ell}$ is evaluated by using eq.~(\ref{Eq:Delta}) and the expectation values 
of $\langle n_{i\ell} \rangle$ in the NM state. 

Finally, we find out possible phases by numerically solving the above self-consistent equations. 
To single out the stablest state, we need to estimate the energy lowering (stabilization energy) 
relative to the energy of the NM state: 
\begin{equation}
\Delta E \equiv \langle H_{\rm HF} \rangle_{\rm NM} - \langle H_{\rm HF} \rangle. 
\end{equation}
The state maximizing $\Delta E$ is the ground state. 
Figure~\ref{fig:energies} depicts the energy lowering of the ordered ground states. 

\begin{figure}[htbp]
\begin{center}
\includegraphics[width=0.70\columnwidth]{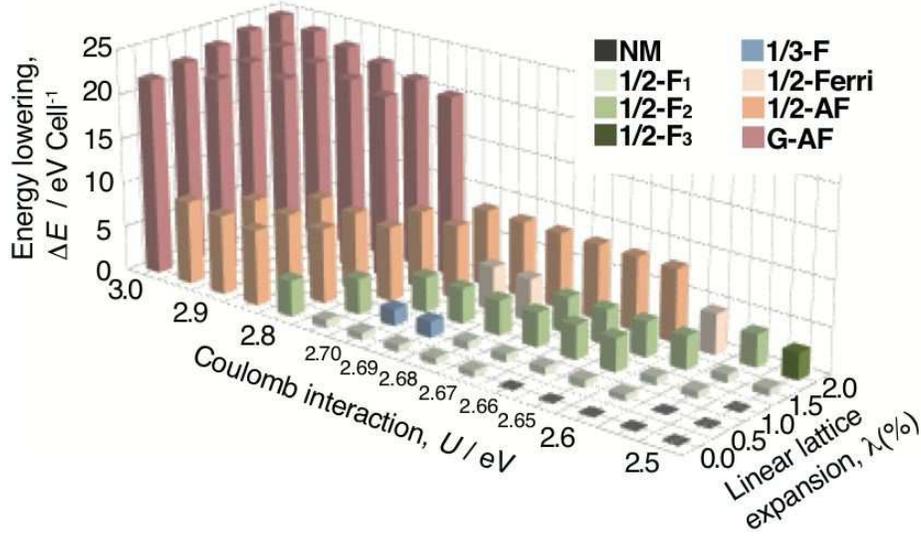}
\end{center}
\caption{
(Color online) Energy lowering in the phase diagram.}
\label{fig:energies}
\end{figure}
%

\subsection{1/2-F$_1$ and 1/2-F$_2$ states}
We compare these two states to the magnetic two states obtained in the Korotin {\it et al.}'s theory~\cite{Korotin_1996}. 
On the one hand, our calculated results are similar to their results in the following points, as partly mentioned in the main text. These two states accommodate 7 electrons, not 6; are different in the amount of $t_{2g}$-to-$e_g$ electron transfer; and are different in the splitting manner of $t_{2g}$ orbital, doublet-singlet or singlet-doublet [Table I in the main text]. 
%
On the other hand, our results are distinct from their results as follows. First, their ordering is based on the G-type antiferromagnetic or uniform ferromagnetic states. Second, their spin-moment values are as large as 2.1 $\mu_{\rm B}$ and 3.2 $\mu_{\rm B}$, which were, hence, identified as the atomic IS and HS states, respectively.  
In contrast, the ordering we found is represented by the modulated wave vectors, which are based on the recent observations for film LaCoO$_3$~\cite{Fujioka_2015}. Furthermore, the spin-moment values are as small as 1.2 $\mu_{\rm B}$ and 2.5 $\mu_{\rm B}$ at the most spin-polarized sites in the 1/2-F$_1$ and 1/2-F$_2$ states, respectively. The former corresponds to the small $t_{2g}$-to-$e_g$ transfer (0.6 electrons), which is supported by the Compton scattering experiments~\cite{Kobayashi_2015}.

These differences probably arise from the difference of adopted $U$ values. They theoretically assumed the effective $U$ value of 7.8 eV~\cite{Korotin_1996}, whereas we selected the value of $U \approx 2.65$ eV by searching the nonmagnetic-magnetic boundary in the ($\lambda$, $U$) parameter space. We also note that the large $U$ value generates the Korotin-like results in our framework of calculations, too. In fact, we obtain the same value of 3.2 $\mu_{\rm B}$ at each Co site for the G-AF state at $U=3$ eV. 

\subsection{Spin-charge order}
\label{SM:Others}
The spin order has been considered up to this point.  In theory, the spin order can be coupled with the charge sector. In Fig.~\ref{fig:spincharge}, we show all the calculated spin-charge structures including the 1/2-F$_1$ and 1/2-F$_2$ states. The large and small spins are accompanied by the relatively positive and negative charges, respectively. 
The amplitude of charge deviation is typically 0.2$e$--0.4$e$, where $e$ denotes the elementary electric charge. 

The existence of charge order has been very recently detected in thin-film LaCoO$_3$ by resonant X-ray scattering~\cite{Sterbinsky_2018}. In bulk LaCoO$_3$, too, the spontaneous dynamical spin-charge deviation has been thus far suggested both experimentally and theoretically~\cite{Heikes_1964, Iguchi_1996, Haverkort_2006, Krapek_2012, Karolak_2015}. In particular, a recent study using dynamical mean-field theory showed that non-local dynamical spin-charge fluctuations should be very strong in bulk LaCoO$_3$ due to the proximity to the thin-film instability of spin-charge order~\cite{Karolak_2015}. 
This theoretical prediction coincides with our observation of the collective dynamical spin-state excitations as the seed of the thin-film structure. 

\begin{figure}[htbp]
\begin{center}
\includegraphics[width=0.7\linewidth, keepaspectratio]{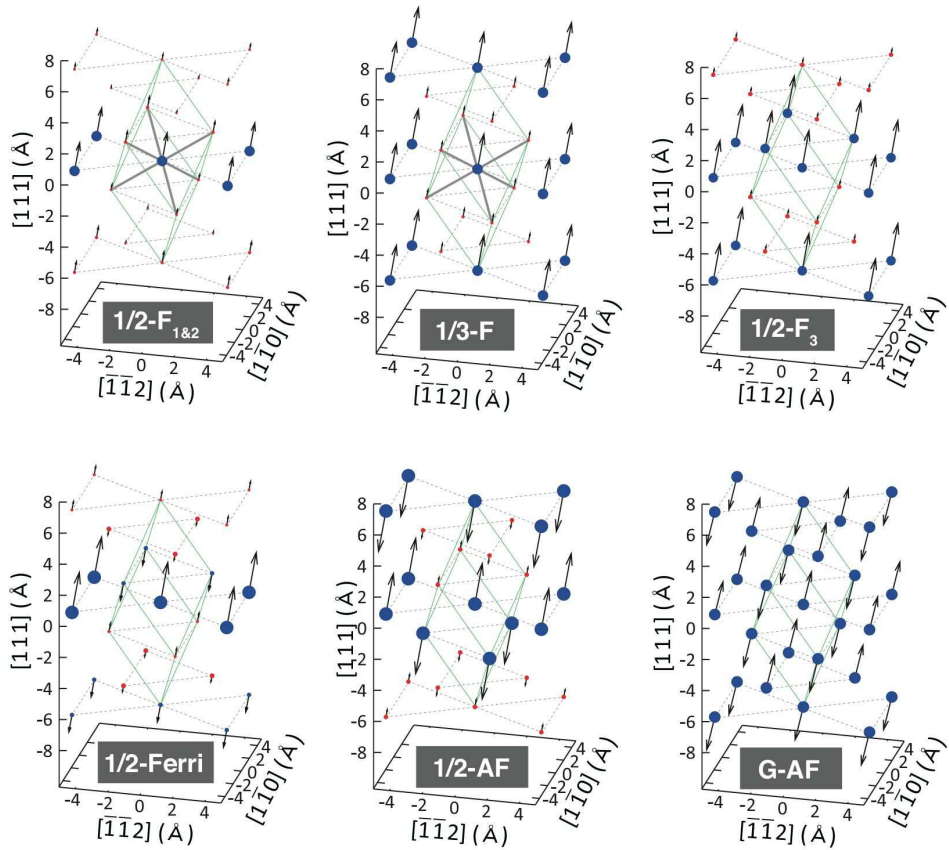}
\end{center}
\caption{\label{fig:spincharge} (Color online)
Real-space spin-charge structures. The green rhombohedron represents the cell, while the thick gray lines show the seven-Co-site unit. The arrows denote the spin moments. Blue and red circles represent the decrease and increase in the electron number relative to the NM state, respectively, and the radii of circles denote its magnitude. The dotted triangular planes indicate the (111) planes. 
The Co-1 and Co-2 sites are inequivalent in the 1/2-F$_1$, 1/2-F$_2$, and 1/2-Ferri states and equivalent in the 1/2-F$_3$, 1/3-F, 1/2-AF, and G-AF states. 
}
\end{figure}

\section{MT range}
\label{SM:MT}
%
Figure~\ref{fig:model} shows the temperature dependence of $\Delta Q_{1}$ and $\Delta Q_{2}$, which are defined as the HWHMs of diffuse scattering and squared magnetic form factor, respectively. Overall, they exhibit no temperature dependence, as described in the main text. 
However, precisely, both the $\Delta Q$s slightly tend to narrow with increasing temperature. This suggests that both the growing of spatial correlation and spatial delocalization are slightly enhanced with increasing the temperature. Therefore, the notion that the $d$-electrons itinerantly support the seven-Co-site collectivity will be verified. 
Furthermore, the progress of delocalization is naturally understood as the precursor phenomenon toward $T_{\rm IM}$, which is akin to Phelan {\it et al.}'s idea of the nanomagnetic droplet of metallicity~\cite{Phelan_2006_a} and experimentally evidences their idea, though their discussion was based on the $d^6$ IS scheme~\cite{Podlesnyak_2006}. 


%
\begin{figure}[htbp]
\begin{center}
\includegraphics[width=0.25\linewidth, keepaspectratio]{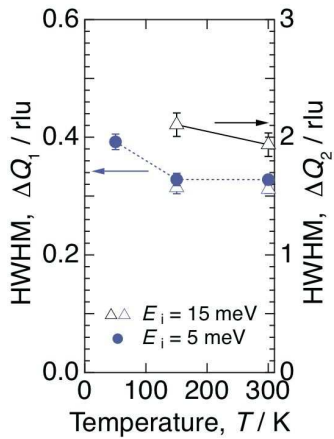}
\end{center}
\caption{\label{fig:model} (Color online)
Temperature dependence of two HWHMs.
}
\end{figure}

We have taken the red dotted arrow shown in Fig.~3(b) (the main text) as the path of thermal change in bulk LaCoO$_3$. However, more precisely, not only $\lambda$ but also $U$, $J_{\rm H}$, and $t$ {\it et al.} thermally changes~\cite{Tomiyasu_2017, Karolak_2015}. Therefore, the rather wide area around the arrow might be swept. As its effect, for example, 1/3-F might be possible as the first magnetic state as well as 1/2-F. In this 1/3-F case, too, the similar seven-Co-site partial structure consistently inheres [the upper central panel in Fig.~\ref{fig:spincharge}], indicating that all the preceding correspondence between theory and experiment is maintained. 

The first spin crossover is coupled to the following lattice anomalies; as the temperature increases, significant softening is observed in the transverse optic phonons around 10 meV, which bend the Co--O--Co angle~\cite{Ishikawa_2004, Kobayashi_2005}, and this angle also structurally increases toward 180$^{\circ}$~\cite{Durand_2013}. The latter increase is contrary to the decrease expected from the tolerance factor of perovskite structure, suggesting the existence of an unresolved electronic origin. Meanwhile, the present seven-Co-site correlation is expected to favor the 180$^{\circ}$ Co--O--Co angle to sustain the $d$--$p$--$d$ orbital hybridization, which will therefore cause these lattice anomalies.  

From the IS aspect, the Jahn-Teller lattice instability has been thus far searched for. The strong experimental evidence of IS nature is the anomalous softening of the Jahn-Teller phonon mode around the high energy of 580 cm$^{-1}$ ($\approx 71$ meV $\approx 830$ K in the unit of Boltzmann constant)~\cite{Ishikawa_2004}. In contrast, the Co--O length does not appear to significantly split in the instantaneous local structure experiments~\cite{Sundaram_2009}. Furthermore, a recent nuclear magnetic resonance study reports that the local lattice symmetry is preserved in the entire MT range~\cite{Shimizu_2017}. These intricate facts could be resolved as follows: because the Jahn-Teller high-energy phonons are not thermally activated in the MT range, the Jahn-Teller lattice instability does not structurally appear so much and is explicitly observed only as the softening and/or damping occurring when forcedly vibrating the lattice. 
To understand the lattice dynamics completely, future investigations of the full phonon dispersion in the wide $Q$, $E$, and temperature ranges may be beneficial. 

Thus, as explained above, the seven-Co-site unit of spin-state excitations is considered the generator of spin-charge-orbital-lattice multi-functions occurring in LaCoO$_3$.

%
\section{HT range}
\label{SM:HT}
In the HT range above $T_{\rm IM}$, the intensity of neutron magnetic diffuse scattering almost disappears~\cite{Asai_1994}, indicating that the correlation effect that characterizes the MT range becomes weak. Meanwhile, the theoretically corresponding 1/2-F$_2$ state exhibits the IS-like spin moment value of 2.5 $\mu_{\rm B}$ with the $t_{2g}$-to-$e_g$ transfers of 1.2 electrons at the most spin-polarized Co site [Table I in the main text]. Furthermore, the aforementioned Jahn-Teller phonon modes also begin to thermally activate. Therefore, the HT range appears IS-like. 
However, the $d$-electron configuration can be roughly regarded as the mid from $d^7$-NM to $d^6$-HS states ($t_{2g}^{4}e_{g}^{2}$) when focusing on the majority orbitals [Table I in the main text]. In addition, the spatial spin correlation slightly persists~\cite{Asai_1994}, which might slightly generate the HS nature. In fact, the summation of seven-Co-site spins is theoretically 5.3 $\mu_{\rm B}$ in the 1/2-F$_2$, which is HS-like rather than IS-like. 
Thus, the HT range is also expected to exhibit the IS and HS dual nature in a manner different from the MT range. This expectation is consistent with the controversial interpretations of magnetic susceptibility data in the HT range~\cite{Jirak_2008}.

\section{Thin film}
\label{SM:film}
Our theory captures the following experimental results for thin-film LaCoO$_3$, as written in the main text; the nonmagnetic-magnetic boundary exists around $\lambda = 0.5${\%} and the modulated order is proximate to the NM~\cite{Fuchs_2007, Fujioka_2015}. 
In addition, the 1/2-F$_1$ described by $\vec{Q} = \vec{Q}_{0}$ and $\vec{Q}_{1/2}$ is theoretically obtained at $\lambda = 1.0${\%}, which is also consistent with the experimental observation of the same $\vec{Q}_{1/2}$ for the $\lambda = 1.0${\%} film~\cite{Fujioka_2015}.

On the other hand, our theory does not exactly reproduce the experimentally observed $\vec{Q} = (1/6,1/6,1/6)$ that appears only in the $\lambda = 0.5${\%} film epitaxially grown on the (111) substrate~\cite{Fujioka_2015}. In the plausible 1/3-F, the Co-1 and Co-2 sites are equivalent, as mentioned in Sect.~\ref{SM:Others}, corresponding to $\vec{Q} = (1/3,1/3,1/3)$, not experimentally observed $(1/6,1/6,1/6)$. However, we certainly found a self-consistent solution with inequivalent Co-1 and Co-2 sites, corresponding to the $(1/6,1/6,1/6)$ ordering, with the slightly higher total energy (1.1 eV per cell) relative to the 1/3-F. Thus, the amplitude-modulated ferromagnetic family is energetically proximate, suggesting that the calculated results are finely tuned to fit with the experimental results by adding a small perturbation term. For example, the complex anisotropic lattice strain depending on the substrate orientation has not been considered this time. Thus, future theoretical and experimental investigations of the relationship between the spatial spin-state order and the anisotropic epitaxial lattice strain type would further clarify the thin-film ferromagnetism. 

\section{Hole-doped system}
Our knowledge obtained for LaCoO$_3$ also gives certain explanation about the hole-doped system, La(Sr)CoO$_3$. The hole doping does not expand the lattice so much~\cite{Kriener_2004} but is expected to promote the spin-charge order through the charge channel. This expectation is consistent with the following experimental facts. First, in lightly hole-doped La(Sr)CoO$_3$, the similar spin-state seven-Co-site polaron (or heptamer magnet) is stabilized as the ground state in the low-temperature range~\cite{Podlesnyak_2008, Podlesnyak_2011}. Second, further hole doping grows the heptamers to the short-range order described by $\vec{Q}_{0}$ and $\vec{Q}_{1/2}$ and generates the insulator-to-metal transition with keeping the same two $\vec{Q}$s~\cite{Phelan_2006_b}. The $\vec{Q}_{0}$ and $\vec{Q}_{1/2}$ vectors are identical to those that describe the 1/2-F$_1$ and 1/2-F$_2$ states. Thus, the present seven-Co-site spin-state excitation unit, characterizing the thermal spin crossover in bulk LaCoO$_3$, is considered to connect the polarons and order extending from insulating to metallic states in the hole-doped system as well as the thin-film ferromagnetic order as their common origin. 
%

%
%

\end{widetext}

\bibliography{LaCoO3_5_arXiv}

\end{document}